# Metal/SrTiO$_3$ two-dimensional electron gases for spin-to-charge conversion

Luis M. Vicente-Arche[1*], Srijani Mallik[1*], Maxen Cosset-Cheneau[2], Paul Noël[2§], Diogo Vaz[1♣], Felix Trier[1♥], Tanay A. Gosavi[3], Chia-Ching Lin[3], Dmitri E. Nikonov[3], Ian A. Young[3], Anke Sander[1], Agnès Barthélémy[1], Jean-Philippe Attané[2], Laurent Vila[2] and Manuel Bibes[1]

[1]Unité Mixte de Physique, CNRS/Thales, Université Paris-Saclay, 91767 Palaiseau, France
[2]Université Grenoble Alpes, CEA, CNRS, Grenoble, INP, IRIG-Spintec, Grenoble, France
[3] Components Research, Intel Corp., Hillsboro, OR 97124, USA

**SrTiO$_3$-based two-dimensional electron gases (2DEGs) can be formed through the deposition of epitaxial oxides like LaAlO$_3$ or of reactive metals such as Al. Such 2DEGs possess a finite Rashba spin-orbit coupling that has recently been harnessed to interconvert charge and spin currents through the direct and inverse Edelstein and spin Hall effects. Here we compare the formation and properties of 2DEGs generated in SrTiO$_3$ by the growth of Al, Ta and Y ultrathin films by magnetron sputtering. By combining *in situ* and *ex situ* X-ray photoelectron spectroscopy (XPS) we gain insight into the reduction of the SrTiO$_3$ and the appearance of Ti$^{3+}$ states associated with 2DEG formation, its reoxidation by exposure to the air, and the transformation of the metal into its binary oxides. We extract the carrier densities through magnetotransport and compare them with the XPS data. Finally, working with samples covered by an extra layer of NiFe, we perform spin-pumping ferromagnetic resonance experiments and investigate spin-charge conversion as a function of gate voltage. We identify trends in the data across the different sample systems and discuss them as a function of the carrier density and the transparency of the metal oxide tunnel barrier.**

* these authors contributed equally to this work
§ now at: Dept. of Materials, ETH Zürich, Hönggerbergring 64, 8093 Zürich, Switzerland
♣ now at: CIC nanoGUNE BRTA, Tolosa Hiribidea, 76, 20018 Donostia - San Sebastian, Spain
♥ now at: Center for Quantum Devices, Niels Bohr Institute, University of Copenhagen, Universitetsparken 5, 2100 Copenhagen, Denmark

1. **Introduction**

Since their discovery in 2004 [1], two-dimensional electron gases (2DEGs) based on SrTiO$_3$ (STO) have been shown to possess a wide array of fascinating properties [2]. They are superconducting below about 200 mK [3], their transport response is very sensitive to gate voltages [4], they harbour a sizeable Rashba spin-orbit coupling (R-SOC) [5,6], and even signs of magnetic order [7]. While these discoveries challenge our understanding of quantum matter, they also trigger some interest for technological applications. Indeed, STO 2DEGs have been functionalized as transistors into integrated circuits [8], as rewritable photodetectors [9] or, more recently, as the read-out unit [10] of future spin-based logic devices coined MESO (for magneto-electric spin-orbit) proposed by Intel Corporation [11]. In this latter example, the R-SOC of the 2DEG is harnessed to efficiently convert spin currents into a charge currents in a bipolar way, through the inverse Edelstein effect [12]. This spin charge conversion was originally demonstrated in STO 2DEGs classically formed by the epitaxial growth of LaAlO$_3$ at high temperature [13,14], but later was also reported in STO 2DEGs generated by the simple sputter deposition of Al at room temperature [10]. This simpler and more integrable synthesis process, combined with the larger conversion efficiency and its survival at room temperature are very appealing for applications and prompted us to investigate metal/STO 2DEGs in greater detail.

In this paper, we report the synthesis of 2DEGs in STO by depositing various reactive metals by magneton sputtering [15]. We selected Al, Ta and Y, which can all be grown as smooth ultrathin films and whose oxide formation enthalpies satisfy the thermodynamic criteria to reduce the SrTiO$_3$ ($\Delta H_{f,ox}$ < -250 kJ/mol O) [16–18]. Using an X-ray photoelectron spectroscopy (XPS) chamber, connected in ultra-high vacuum (UHV) to our sputtering chamber, we could monitor the formation of the 2DEG through the appearance of Ti$^{3+}$ states, as well as its degradation after the samples were exposed to the air. We confirmed the presence of the 2DEG through temperature dependence resistance measurements and Hall experiments from which we deduced the carrier density in a two-band model. We found that both the Ti$^{3+}$ concentration and the carrier density tend to increase with the metal thickness deposited, with subtle differences between the three systems. Finally, we prepared samples in which the reactive metal was covered by a Ni$_{80}$Fe$_{20}$ (permalloy, NiFe) layer so as to perform spin-pumping experiments by ferromagnetic resonance (SP-FMR). We measured the charge current generated transverse to the injected spin current due to the inverse Edelstein effect as a function of the gate voltage for samples based on Al, Ta and Y, having different ungated carrier density values. We finally compared the gate dependences across various sample types to draw conclusions on the role of the complex multi-orbital band structure on the spin-charge conversion process.

2. **Experimental Details**

**Preparation of STO substrates:** Single-crystal STO (001) substrates (5 mm × 5 mm × 0.5 mm, one-side polished with miscut angles < 0.1°) were purchased from SurfaceNet GmbH. The as-received substrates, with a mixed surface termination (SrO-$TiO_2$), were cleaned by sonicating in deionized water, acetone and isopropanol and subsequently dried with nitrogen. This process was repeated until the cleanliness of the surface was confirmed by AFM. The cleaned substrates were inserted in a UHV system for metal deposition by magnetron sputtering and in-situ XPS measurements.

**Metal deposition:** The metal deposition was performed in a commercial dc magnetron sputtering system PLASSYS MP450S with a base pressure of 9 x $10^{-8}$ mbar. The different deposition rates were derived by means of x-ray reflectometry (XRR) on thicker samples grown under the same conditions. All metal depositions were carried out on different STO substrates at room temperature. Ar gas flow and the current intensity were fixed to 5.2 sccm and 30 mA, respectively. The pressure of the chamber during depositions was 5.3 ± 0.2 x $10^{-4}$ mbar. The plasma power varied slightly depending on which metal was being deposited: 10 W for Al, 9 W for Ta and 8 W for Y. The NiFe was also deposited by dc magnetron sputtering, and capped with 1.5 nm of Al, that transformed into $AlO_x$ after exposure to air.

**XPS measurements:** The XPS measurements were performed at room temperature using an Omicron NanoTechnology GmbH system with a base pressure of 5 x $10^{-10}$ mbar, using a Mg $K_\alpha$ source ($hv$ = 1253.6 eV) operating at 20 mA and 15 kV. The spectra were obtained at a pass energy of 20 eV. In-situ XPS measurements were performed before and immediately after the deposition of Al, Ta or Y on different samples. Ex-situ XPS measurements were also carried out on the same samples after they were exposed to air. The fitting of the spectra was carried out using CasaXPS (CasaSoftware Ltd.).

**Magnetotransport properties measurements:** The samples were measured with a Dynacool system from Quantum Design after bonding with Al wires in Van der Pauw configuration. During the transport measurements of AlO$_x$/NiFe/metal//STO samples, the NiFe and 2DEG signal were probed in parallel [19]. Both contributions were disentangled considering the following: for the longitudinal configuration, $R_{xx}$, two resistances in parallel were measured so that the resistance of the 2DEG alone is given by

$$R_{2DEG} = (R_M \times R_{Total}) / (R_M - R_{Total})$$

For the transverse configuration, $R_{xy}$ (Hall resistance), besides $R_M$ and $R_{2DEG}$ the Hall voltages generated in each layer must be also considered. These circuits can be simplified using Millman's theorem [20] so that the Hall resistance of the 2DEG alone $R_{H,2DEG}$ is given by

$$R_{H,2DEG} = R_{H,Total} \times ((R_M / R_{Total}) + 1)^2 - R_{H,M} \times (R_{2DEG} / R_M)^2$$

The 2DEG contribution was then fitted with a standard two-band model in order to extract carrier densities and mobilities.

**Spin pumping:** The spin-pumping experiments were carried out using a Bruker ESP300E X-band CW spectrometer at 9.68 GHz, with a loop-gap Bruker ER 4118X-MS5 cavity, and using a microwave power of 5 mW or less to remain in the linear regime and avoid thermal effects [21]. The DC voltage generated transverse to the magnetization was measured using a Keithley 2182A nanovoltmeter. The gate voltage was applied using a Keithley 2400 sourcemeter. The sample was initialized by sweeping the back-gate voltage from +200 V to −200 V, and then back to +200 V, to avoid any hysteretic behavior. The measurement was then performed for different gate voltages, from +200V to -200V. The measured signals were observed to be linear with rf power up to 5 mW with no indications of heating [21].

3. Results
    a. XPS

Prior to any deposition, XPS was performed individually on all bare STO substrates used to prepare samples for this work. All showed similar spectra for the Ti 2p core-levels, an example of which is shown in the inset of Fig. 1A. During the fitting, adventitious carbon was used as a charge reference to obtain the Ti$^{4+}$ $2p_{3/2}$ peak position ~458.6 eV that is consistent with previously reported values for STO [22–25]. The fits of the spectra suggest the presence of $Ti^{4+}$ only, as expected for as-received insulating STO single crystals.

The Ti 2p core-level spectra, collected in-situ after depositing different metal thicknesses, are shown in Fig.1 (Al (Fig 1.A - C), Ta (Fig 1.D - F) and Y (Fig 1.G - I)). Signs of the reduction of STO are visible from the spectra as the lower binding energy peaks of $Ti^{4+}$ ($Ti^{3+}$ and $Ti^{2+}$) appear upon deposition of the metal. Further, the $Ti^{3+}$ and $Ti^{2+}$ peak areas become larger with increasing metal thicknesses.

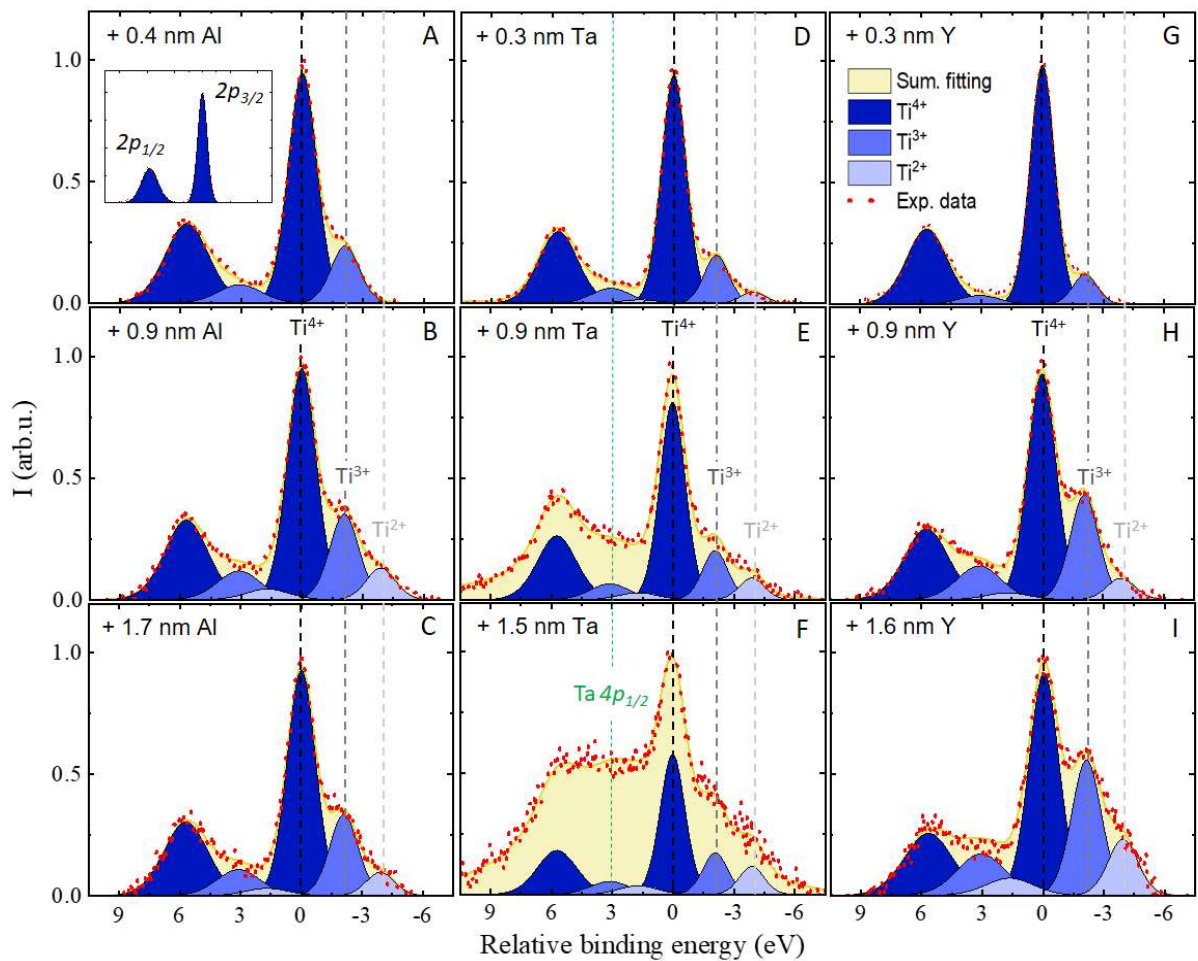

**Fig.1**: Ti 2p core-level spectra obtained in-situ when different thicknesses of: aluminium (**A - C**), tantalum (**D - F**) and yttrium (**G - I**) are deposited on STO. Ti 2p spectra for the bare STO substrate is shown as an inset in **A**. All the spectra have been normalized to the maximum intensity and the binding energy values are relative to the $Ti^{4+}\ 2p_{3/2}$ peak position. The colour code is the same for all figures and it is shown in **G**. Note that, for the Ta case, the Ti

2p core-level energy region overlaps with the energy range of Ta $4p_{1/2}$. Therefore, for better visualization the fitted Ta peaks have been removed from the spectra.

The Ti 2p core-level spectra collected after exposing the same samples to air are shown in Fig.2 (Al (Fig2.A - C), Ta (Fig2.D - F) and Y (Fig2.G - I)).

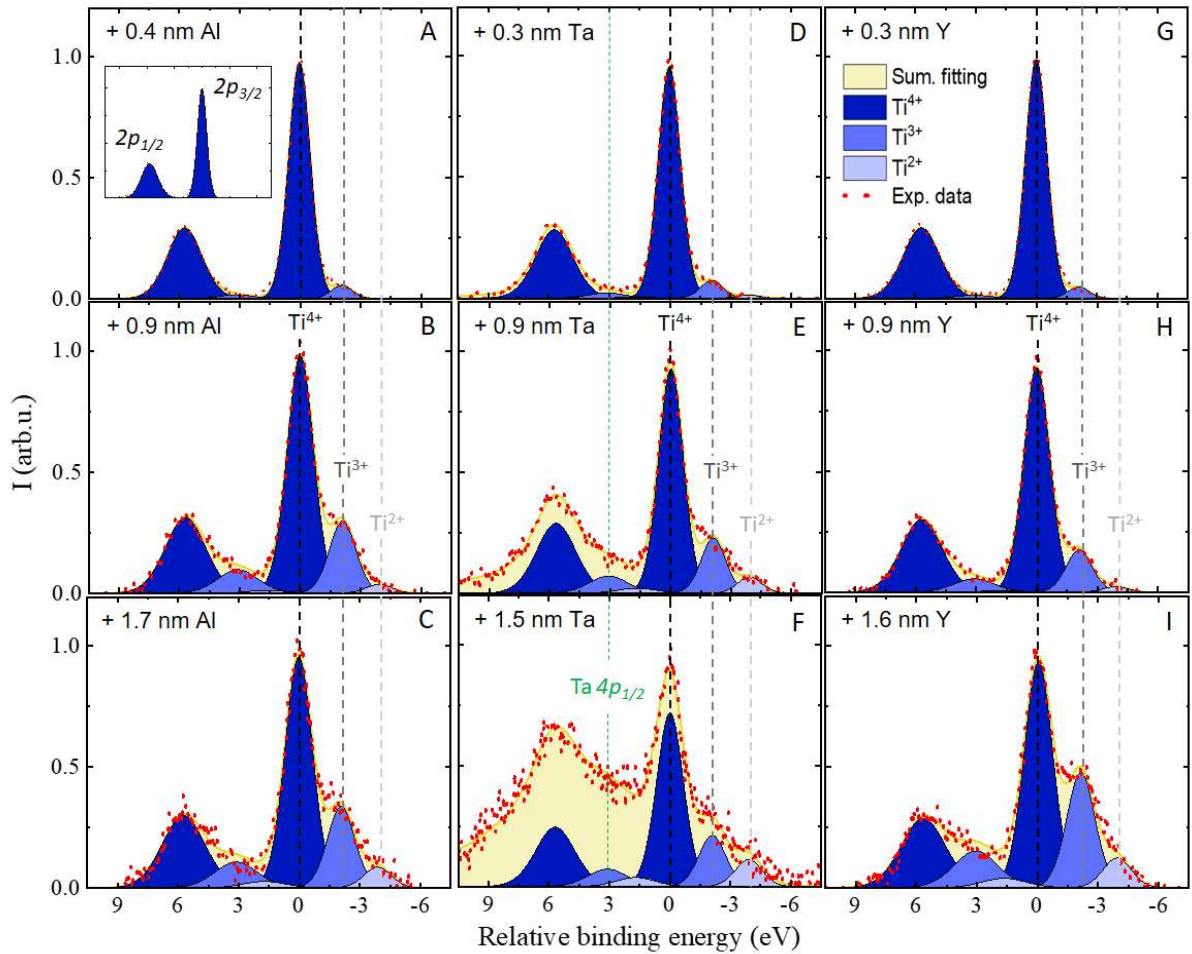

**Fig.2**: Ti 2p core-level spectra obtained ex-situ when different thicknesses of aluminium (**A - C**), tantalum (**D - F**) and yttrium (**G - I**) are deposited on STO. All the spectra have been normalized to the maximum intensity and the binding energy values are relative to the $Ti^{4+}$ $2p_{3/2}$ peak position. The colour code is the same for all figures and it is shown in **G**.

For the Ta case, the Ti 2p core-level energy region overlaps with the energy range of Ta $4p_{1/2}$. Thus, it was necessary to consider its contribution to determine the Ti valence states. In this case, we analysed a wider energy region which included the Ta $4p_{3/2}$ core-level in order to constrain the contribution of Ta $4p_{1/2}$ during the analysis.

The contribution of the reduced titanium was determined by means of an active fitting process during which all possible valence states were considered. The peak area ratio Ti $2p_{1/2}$ : Ti $2p_{3/2}$ was fixed to 0.5 and fitted with a Gaussian-Lorentzian (70:30) peak line shape. Finally, we achieved the best agreement for all the spectra when the following constrains were applied:

|   | Peak | ΔE ($2p_{1/2} - 2p_{3/2}$) | FWHM | B.E. |
|---|---|---|---|---|
| A | $Ti^{4+}$ $2p_{1/2}$ | 5.7 eV | 2.5 ± 0.1 eV | - |
| B | $Ti^{4+}$ $2p_{3/2}$ |  | 1.7 ± 0.1 eV | 458.6 eV |
| C | $Ti^{3+}$ $2p_{1/2}$ | 5.2 eV | A*1 | - |
| D | $Ti^{3+}$ $2p_{3/2}$ |  | B*1 | 456.5 eV |
| E | $Ti^{2+}$ $2p_{1/2}$ | 5.6 eV | A*1 | - |
| F | $Ti^{2+}$ $2p_{3/2}$ |  | B*1 | 454.7 eV |

**Table 1.** Fitting parameters used for the XPS Ti 2p spectra displayed in Fig. 1 and Fig.2. FWHM stands for full width at half-maximum and B.E. for binding energy.

After the analysis, we confirmed that titanium reduces differently depending on the metal deposited. For a better visualization, the evolution of Ti valence states is plotted (Fig. 3) as a function of the metal thicknesses for all the samples measured in-situ (open circles) as well as after being exposed to air (filled circles). Comparing the relative concentration of $Ti^{3+}$ present in the STO substrates before exposing the samples to air (open circles), it can be inferred that most of the reduction of STO occurs for the first few angstroms deposited (0.5 nm) and is highest for Y, than for Al and for Ta.

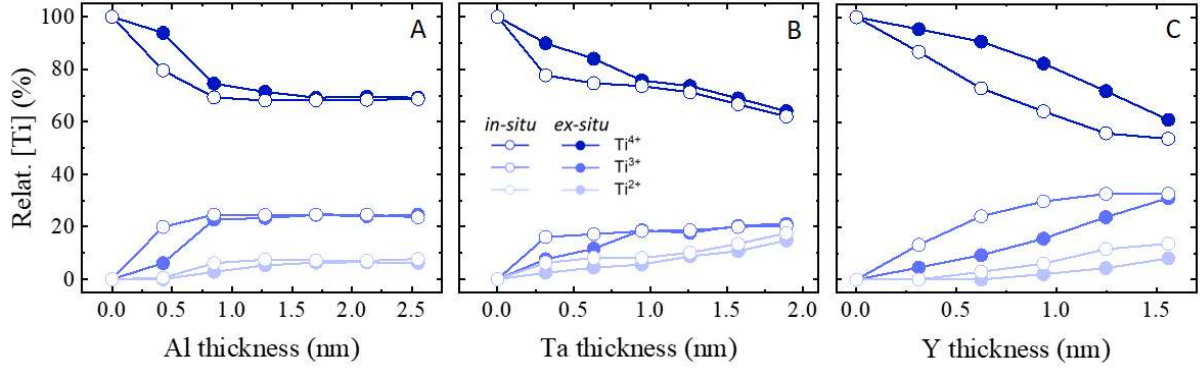

**Fig. 3**: Ti valence states evolution when increasing the thickness of: Al (**A**), Ta (**B**) and Y (**C**). The open circles correspond to the spectra collected in-situ. The filled circles correspond to the spectra collected for the same samples after being exposed to air.

Furthermore, Fig. 3 also reflects the critical thickness needed to preserve the reduced Ti from reoxidation by the air. This critical thickness can be defined as point where the in-situ and ex-situ curves for the $Ti^{4+}$ valence state meet. For Al and Ta, the critical thickness is around 1 nm and it is greater than 1.5 nm for Y. This observation can be well correlated with the oxygen diffusion coefficients of $Y_2O_3$, $Ta_2O_5$, and $Al_2O_3$. It is known that the oxygen diffusion coefficient for $Ta_2O_5$, and $Al_2O_3$ are of similar order of magnitude whereas it is higher for $Y_2O_3$ [26]. Therefore, even if Y shows the highest reduction of STO, it is not a good candidate to protect the reduced metal/STO interface (and thus the 2DEG) in comparison with Al and Ta.

In addition, we also studied the evolution of the metal valence states after each deposition. For the first depositions (~1 nm), it is clearly visible that the metals get oxidized at the STO interface. Further, by increasing the metal thickness we observe the contribution of non-oxidized metal along with its oxide states. As an example, Fig. 4 shows ex-situ spectra having both non-oxidized metal and corresponding metal oxides contributions for specific thicknesses of the three metals.

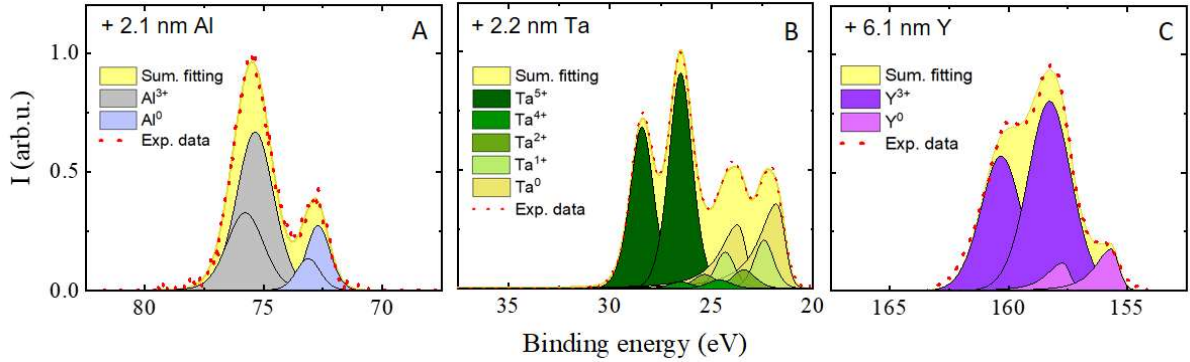

**Fig. 4** XPS spectra obtained after the deposition of 2.1 nm of aluminium (**A**), 2.2 nm of tantalum (**B**) and 6.1nm of yttrium (**C**). It shows that aluminium and yttrium are mostly oxidized with a valence state 3+ which could correspond to $Al_2O_3$ and $Y_2O_3$ respectively, whereas in the case of tantalum oxide several valence states (1+, 2+, 4+, 5+) are present.

The relative fraction of oxides and non-oxidized metal contributions obtained ex-situ for different materials are shown in Fig. 5 as a function of metal thicknesses. By increasing the metal thickness, we observe the non-oxidized metal contribution for Al and Y along with their corresponding oxide valence state ($Al^{3+}$, $Y^{3+}$). In contrast, in the case of Ta, first we observe the appearance of reduced valence states of Ta related to the formation of intermediate oxides ($Ta^{4+}$, $Ta^{2+}$, $Ta^{1+}$) which exhibit maxima depending on the thickness of the deposited Ta layer. By further increasing the thickness we observe the contribution from metallic Ta. It should be noted that the appearance of a non-oxidized metal is observed in ex-situ XPS spectra above a minimum thickness of the metal layers. Ideally, this contribution appears when no further oxidation of the metal is possible between the oxide layers formed by the oxidation from STO and from the air.

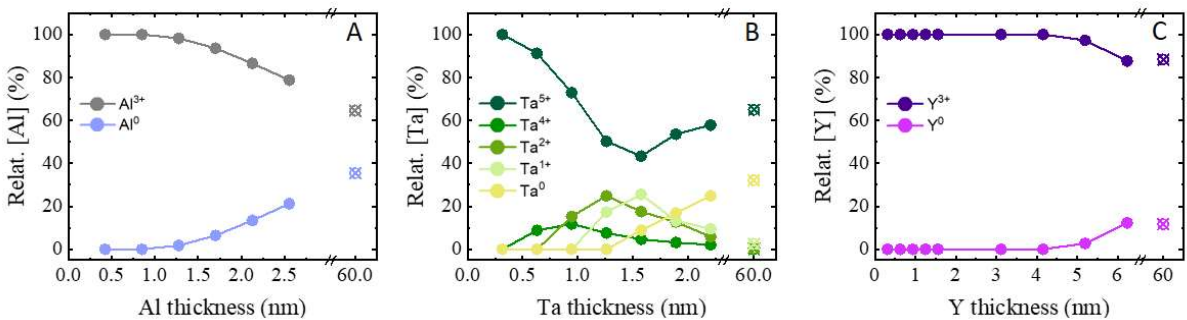

**Fig. 5:** Relative fractions of oxide and metals for aluminium (A), tantalum (B), and yttrium (C) as a function of the metal thickness.

It can be observed in Fig. 5 that the minimum thickness at which the non-oxidized metal appears is similar for the Al and Ta cases (~1.5 nm), while in the case of Y it is notably higher (~5 nm). This can be explained by considering two phenomena happening simultaneously. The first is the stronger oxidation of Y upon reducing STO as observed in Fig. 3. The second is the higher oxygen diffusion coefficient [26–28] and thicker metal oxide layer thickness for Y when exposed to air. This is also confirmed by the metal oxide layer thicknesses ($t_{ox}$) obtained by means of XRR on thicker samples which are $t_{ox,Y_2O_3}$≈3.8 nm, $t_{ox,Al_2O_3}$≈1.8 nm, $t_{ox,Ta_2O_5}$ ≈1.6 nm. Further, by increasing the thickness of the metal ~60 nm, the trends evolve towards the saturation of contributions from the non-oxidized metal and oxides where only the native oxide and non-oxidized metal underneath are detected (as shown by the crossed circle in Fig. 5).

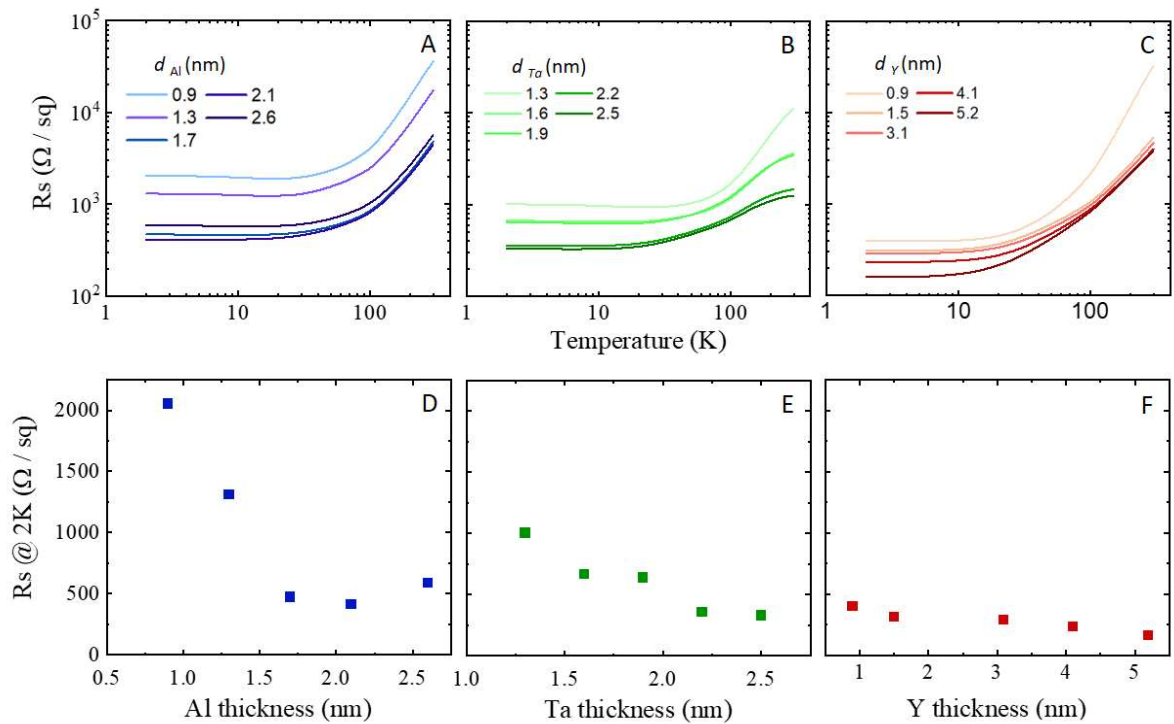

**Fig. 6**: (Top row) Temperature dependence of the sheet resistance of the samples when varying the thickness of the aluminium (**A**), tantalum (**B**) and yttrium (**C**). (Bottom row). Plots for the sheet resistance at T=2K by varying the thickness of the aluminium (**D**), tantalum (**E**) and yttrium (**F**).

### b. Transport measurements in metal//STO

In Fig. 6 we show $R_S$ vs T plots and the sheet resistance at 2 K of samples with different metal thicknesses. We observe a variation of $R_S$ by one order of magnitude, which cannot be ascribed to the metallic capping thus implying the existence of an additional path for conduction related to the formation of a 2DEG when the STO is reduced, consistent with the XPS data. Fig 6(D – F) show the thickness dependence of sheet resistance ($R_S$) at 2K for Al, Ta and Y, respectively. The sheet resistance increases with decreasing metal thickness and the samples become insulating below ~1 nm as the oxide layer is not enough to protect the 2DEG region from re-oxidation.

Fig. 7(A - C) show the non-linear Hall resistance curves at 2K for each system and for different metal thicknesses. The carrier densities and mobilities of the different 2DEGs have been extracted by fitting the Hall data (Fig. 7(A - C)) using a 2-band model and the sheet resistance at 2 K (Fig. 6(D – F)). A general trend is that the carrier density increases with the deposited metal thickness until it saturates, for different values depending on the metal: $n_{Al} \approx 9 \times 10^{13}$ cm$^{-2}$, $n_{Ta} \approx 7 \times 10^{13}$ cm$^{-2}$, $n_Y \approx 4.5 \times 10^{13}$ cm$^{-2}$. In addition, we also observe that the saturation in carrier density occurs at different thicknesses depending on the metal deposited. This occurs when there is no further reduction of STO by increasing the metal thickness. For Al, Ta and Y the saturation of carrier densities occurs at 1.6 nm, 1.2 nm and 1.8 nm, respectively. These thicknesses match well with the critical metal thicknesses obtained from XPS, cf. Fig. 3.

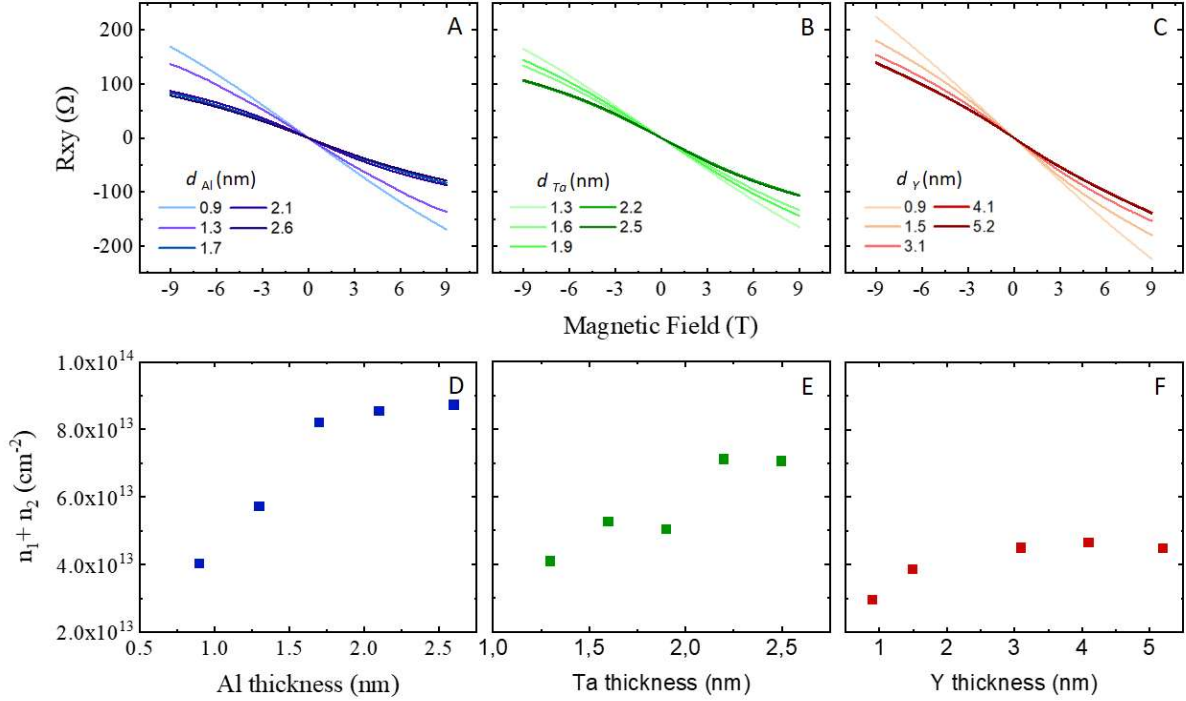

**Fig. 7**: Hall resistance curves for Al (A), Ta (B) and Y (C) and carrier densities of the 2DEG at T = 2 K for Al (D), Ta (E) and Y (F) as a function of metal deposited thickness.

By comparing the carrier densities of each system, it can be observed that Al shows the highest carrier density while Y exhibits the lowest. However, we observed from the XPS data that the reduction of STO was maximum for Y. So, it was expected to have higher carrier densities in Y. Therefore, it has not been possible to correlate the concentration of $Ti^{3+}$ into the STO with the transport properties of the 2DEG hosted at the interface between different oxides. As reflected in this study, a higher concentration of oxygen vacancies in STO does not necessarily lead to higher densities of mobile carriers. The formation of 2DEG are complex process where the ionic and electronic transport through the oxide layer seems to be determinant factor in the final properties of the 2DEG. Also, the role of metal-induced gaps states, defects and fixed charges must be considered.

c. **Transport measurements on NiFe/metal//STO samples**

To perform spin pumping experiments, we prepared samples in which an ultrathin metal layer (0.9-1 nm of Al or Ta or Y on a STO substrate) was covered by a NiFe thin film, and then capped with Al (that fully oxidizes when the samples are taken out of the chamber, and prevents oxidation of the NiFe). Two sets of samples were prepared with either 2 nm (for transport experiments) or 20 nm (for spin-pumping experiments) of NiFe. The results for samples with 0.9-1 nm of metal are shown in Table 2. The carrier concentration increases when going from Ta to Al and to Y. The trend is consistent with that of the $Ti^{3+}$ fraction measured in XPS (extracted from Fig. 3), which indicates that the NiFe prevents the reoxidation of the STO after exposure to air.

| Metal | $Ti^{2+}$ (%) | $Ti^{3+}$ (%) | $Ti^{4+}$ (%) | $n_{tot}$ (cm$^{-2}$) |
|---|---|---|---|---|
| Ta | 9.5 | 18.6 | 73.6 | 6.66 10$^{13}$ |
| Al | 6.1 | 24.6 | 69.3 | 9.22 10$^{13}$ |
| Y | 6.1 | 29.8 | 64.1 | 10.5 10$^{13}$ |

**Table 2.** Relative fraction of Ti species from *in situ* XPS in metal//STO samples with 0.9 nm-1 nm thick metal layer and 2DEG carrier density extracted from Hall data for NiFe(2 nm)/metal(0.9-1 nm)//STO samples.

### d. Spin pumping experiments in NiFe/metal//STO samples

Fig. 8 presents the results of the SP-FMR experiments on three NiFe/metal//STO samples with 1 nm thick metal layers. At the ferromagnetic resonance, a pure spin current is injected from the NiFe into the STO 2DEG with a spin-polarization parallel to the in-plane magnetization. The data shown in Fig. 8 correspond to the transverse voltage generated when ferromagnetic resonance is reached. Angle-dependent measurements (not shown here) discard spin rectification effects and attribute the measured signal to spin charge conversion occurring at the Rashba 2DEG due to the inverse Edelstein effect [21,29]. Panels A, B and C of Fig. 8 display this signal as a function of the back gate voltage and the data present a number of similarities. First, for all three systems the measured voltage is large, corresponding to signals one to two orders of magnitude higher than what is typically found with e.g. Pt or Ta. This is consistent with previous results in STO 2DEGs that showed remarkably high spin-charge conversion efficiencies [10,13,14,30,31]. The figure of merit ($\lambda_{IEE}$) of the spin-charge conversion efficiencies is defined as the ratio between the detected two-dimensional charge current and the injected three-dimensional spin current, and is expected to be proportional to the system's Rashba coefficient and the momentum relaxation time, i.e. $\lambda_{IEE} = \alpha_R \tau / \hbar$. STO 2DEGs exhibit values of $\lambda_{IEE}$ in excess of 10 nm, i.e. higher than in other Rashba systems [12] or topological insulators [32]. Second, the signal strongly varies with the gate voltage, showing several sign changes with negative and positive maxima. This control of the spin-charge conversion constitutes an additional degree of freedom with respect to the spin-orbitronics stacks based on the spin Hall effect. In an earlier paper combining SP-FMR data with angle-resolved photoelectron spectroscopy and theoretical calculations, we showed that these features can be related to the complex multi-orbital band structure of STO 2DEGs [10]. The observed maxima correspond either to the onset of conversion in bands having opposite signs for their spin-charge conversion efficiencies or to specific avoided crossing points, some of which have a topological character [13,33].

The maxima occur at slightly different gate voltage values for Ta, Al and Y samples, which reflects slight differences in the position of the Fermi energy across the band structure as carriers are accumulated or depleted into the 2DEG by the gate voltage. This is expected since the carrier densities in the ungated case are different for each metal, cf. Table 2. Also, the difference in gate voltage positions between the two negative maxima varies between samples, being around 200 V for Al and Y but closer to 300 V for Ta. This may reflect different efficiencies in the electrostatic gating, for instance, due to different built-in electric fields in the 2DEG and/or different dielectric constant values.

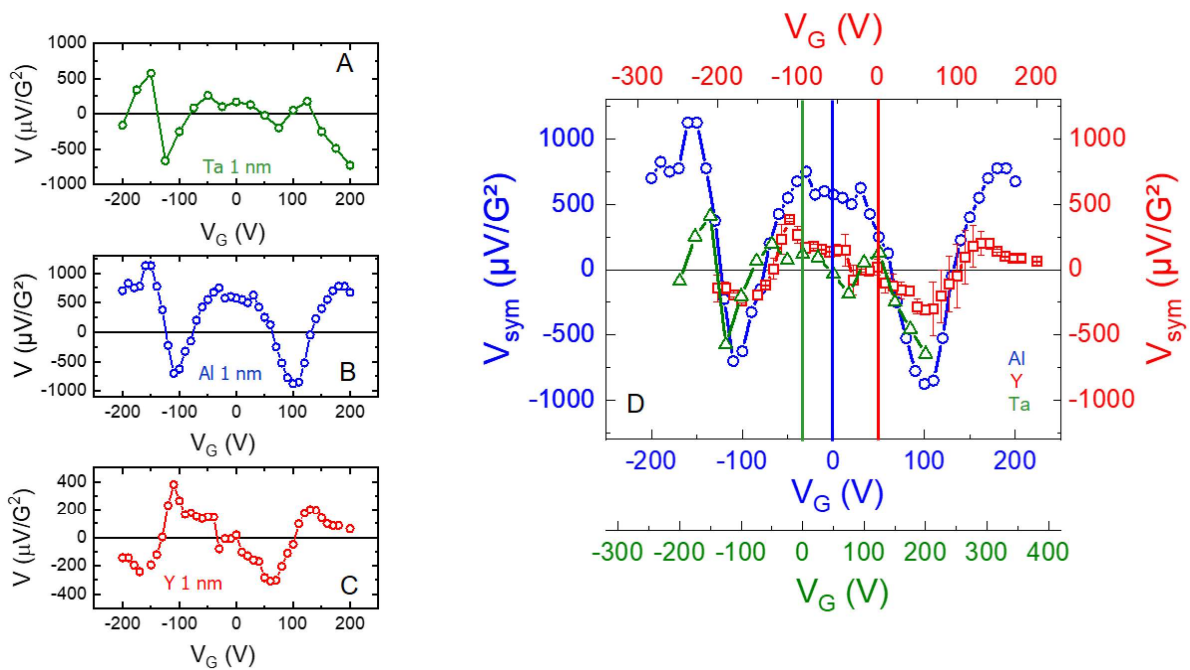

**Fig. 8.** Gate dependence of the transverse voltage measured during SP-FMR experiments for NiFe(20 nm)/metal(1 nm)//STO samples for Ta (A), Al (B) and Y (C). D. Same data replotted by shifting and/or expanding the gate voltage axis for better comparison (see text for details). The vertical lines indicate the zero gate position for the three systems. The data for Al are replotted from Ref. [10].

Fig. 8D replots the data from Figs. 8A, 8B and 8C with the same vertical scale but with the horizontal axes slightly shifted and/or stretched to match the position of the two negative maxima. The shifts are consistent with a Fermi level position at zero gate voltage being lower for Ta than for Al and for Y. That is, with an ungated carrier density lower from Ta than for Al and for Y, in line with the results of Table 2.

While the main features of the spin-charge conversion efficiency as a function of gate voltage can be explained by band structure effects and subtle variations in the carrier density between samples, the absolute amplitude of the conversion differs between Al/STO, Ta/STO and Y/STO 2DEGs. This is a priori surprising if the various maxima observed in the gate dependences correspond to the same points in the band structure for all three systems. In fact, another ingredient is needed: as mentioned above, the conversion efficiency is proportional to the (effective) Rashba coefficient $\alpha_R$ and to the scattering time $\tau$. $\tau$ is not restricted to the 2DEG only, but must also accounts for the possibility for electrons to leak out of the 2DEG through the metal oxide – acting as a tunnel barrier – to quickly scatter in the NiFe (the scattering time in metals is in the fs range, while it is in the ps range in the 2DEG). In fact, Al, Ta, and Y oxide tunnel barriers have different tunneling transparencies, quantified by their resistance area (RA) product and may thus result in various effective $\tau$ values. We can relate the escape time to the RA product through $\tau_{esc} = \frac{RAe^2 m^*}{2\pi\hbar^2}$ with e the electron charge and m* the effective mass and thus calculate the effective scattering time $\tau_{eff}=(\tau_{2DEG}^{-1} + \tau_{esc}^{-1})^{-1}$ with $\tau_{2DEG}$ the scattering time in the 2DEG. While ~1 nm thick AlO$_x$ and TaO$_x$ barriers have relatively high RA values, in the order of 1 MΩ.µm², corresponding to $\tau_{esc}$ in the ns range or longer, YO$_x$ barriers are less resistive with RA values closer to 1 kΩ.µm² [34–36], leading to a $\tau_{esc}$ in the ps range. Since $\tau_{2DEG}$ is also in the ps range, significant leakage is expected to occur in the Y sample, leading to a reduction of the spin-charge conversion efficiency, as observed experimentally.

## 4. Conclusion

In summary, we have investigated the generation of 2DEGs in STO by the deposition of various reactive metals (Al, Ta, Y) for a broad range of metal thicknesses. The 2DEG forms when the metal reacts with the oxygen ions present in the STO, creating oxygen vacancies and releasing electrons. We have monitored this process through XPS and notably associated the appearance of $Ti^{3+}$ with the population of the Ti $t_{2g}$ states. The $Ti^{3+}$ content increases with the metal thickness deposited, saturating more quickly for Al and Ta than for Y. The sheet resistance and carrier density decrease and increase, respectively, with the metal thickness for all three systems, which provides a convenient means to design 2DEG with a specific carrier density, something that is hard to control with $LaAlO_3$/$SrTiO_3$ 2DEGs. We also combined these 2DEGs with ferromagnetic overlayers used to inject a spin current into the 2DEG, and studied the spin-charge conversion process due to the inverse Edelstein effect as a function of the gate voltage. We found very large conversion efficiencies and a complex gate dependence, consistent with earlier results in $LaAlO_3$/$SrTiO_3$ and Al/STO 2DEGs, confirming that the complex multi-orbital band structures drives the spin-charge conversion physics. The tunability of the 2DEG properties, their simple elaboration procedure and their efficient spin charge conversion capability not only make STO 2DEGs appealing for MESO-type devices but should also motivate experiments to probe further charge-spin conversion and spin-orbit torque. More generally, our results expand the research on STO 2DEGs for spintronics and open avenues towards the design of STO 2DEGs based on other reactive metals displaying functional properties [37–39].

**Acknowledgements**

This work received support from the ERC Advanced grant n° 833973 "FRESCO", the QUANTERA project "QUANTOX", the Laboratoire d'Excellence LANEF (ANR-10-LABX-51-01), the ANR project TOPRISE (ANR-16-CE24-0017) and Intel's Science and Technology Center - FEINMAN. F. Trier acknowledges support by research grant VKR023371 (SPINOX) from VILLUM FONDEN.